\documentclass[10pt, twocolumn]{article}
\usepackage[a4paper, total={7in,10in}]{geometry}
\usepackage{titling}
\usepackage{titlesec}
\usepackage[utf8]{inputenc}
\usepackage{amsmath}
\usepackage{amsfonts}
\usepackage{amssymb}
\usepackage{graphicx}
\usepackage{hyperref}
\usepackage{authblk}
\hypersetup{
    colorlinks=true,
    citecolor=red
}  
\titleformat*{\section}{\normalsize \bfseries}

\begin{document}
\title{Effects of plasma turbulence on the nonlinear evolution of magnetic island in tokamak}
\author[1]{Minjun J. Choi \thanks{mjchoi@nfri.re.kr}}
\author[2]{L\'{a}szlo Bard\'{o}czi}
\author[1]{Jae-Min Kwon}
\author[3]{T. S. Hahm}
\author[4]{Hyeon K. Park}
\author[1]{Jayhyun Kim}
\author[1]{Minho Woo}
\author[1]{Byoung-Ho Park}
\author[5]{Gunsu S. Yun}
\author[4]{Eisung Yoon}
\affil[1]{National Fusion Research Institute, Daejeon 34133, Korea, Republic of}
\affil[2]{General Atomics, P.O. Box 85608, San Diego, California 92186-5608, USA}
\affil[3]{Seoul National University, Seoul 08826, Korea, Republic of}
\affil[4]{Ulsan National Institute of Science and Technology, Ulsan 44919, Korea, Republic of}
\affil[5]{Pohang University of Science and Technology, Pohang, Gyungbuk 37673, Korea, Republic of}
\affil[ ]{}
\affil[ ]{\textit {mjchoi@nfri.re.kr}}

\twocolumn[
\begin{@twocolumnfalse}
\maketitle
\begin{abstract}
Magnetic islands (MIs), resulting from a magnetic field reconnection, are ubiquitous structures in magnetized plasmas.
In tokamak plasmas, recent researches suggested that the interaction between the MI and ambient turbulence can be important for the nonlinear MI evolution, but a lack of detailed experimental observations and analyses has prevented further understanding.
Here, we provide comprehensive two-dimensional observations that indicate various effects of the ambient turbulence on the nonlinear MI evolution.
It is shown that the modified plasma turbulence around the MI can lead to either destabilization or stabilization of the MI instability in tokamak plasmas. 
In particular, significantly enhanced turbulence at the X-point of the MI results in a violent disruption through the fast magnetic reconnection and magnetic field stochastization.
\end{abstract}
\vspace{0.5cm}
\end{@twocolumnfalse}
]

Magnetohydrodynamic instabilities accompanied by magnetic island (MI) are a serious concern in tokamak research since they destroy the nested structure of magnetic flux surfaces and lead to degradation of the plasma confinement.
MIs can grow spontaneously through the classical~\cite{Furth:1963hd} and/or neoclassical~\cite{Carrera:1986cca} mechanisms in tokamak plasmas.
When MIs are driven by free energy from the gradient of the equilibrium plasma current, it is known as the classical tearing mode~\cite{Furth:1963hd}.
On the other hand, the neoclassical mechanisms are related to perturbations of the non-inductive current such as the bootstrap current ($J_\mathrm{BS}$).
The $J_\mathrm{BS}$, which is driven by the pressure gradient, can be diminished inside the MI due to the pressure profile flattening.
It occurs when the transport along the radially reconnected field line dominates to equilibrate temperature across the MI.
The loss of $J_\mathrm{BS}$ can reinforce the initial deformation of the flux surfaces, accelerating the magnetic reconnection process.
This is called as the neoclassical tearing mode (NTM)~\cite{Chang:1995una, Haye:2006gh}, driven dominantly by the $J_\mathrm{BS}$ loss over the classical mechanism. 
Observation of the NTM~\cite{Chang:1995una} demonstrates that the transport around the MI is important for the tearing mode stability in tokamak plasmas in addition to the other reconnection physics. 

Recent plasma experiments~\cite{Zhao:2015gra, Rea:2015he, Estrada:2016gz, Bardoczi:2016gj, Bardoczi:2017gq, Choi:2017ez, Chen:2017kb, Jiang:2018fz, Choi:2018uv, Jiang:2019fi, Sun:2018hh} and simulations~\cite{Connor:2009je, Poli:2009ce, Hornsby:2010fh, Ichiguchi:2015hr, Zarzoso:2015fh, Izacard:2016de, Hu:2016kea, Navarro:2017ei, Kwon:2018bi, Fang:2019fz, Ishizawa:2019ky} found that the MI affects the evolution of the ambient broadband fluctuations (or simply referred as `plasma turbulence'~\cite{Yamada:2008fx}). 
Plasma turbulence is significantly altered by the MI itself or a modification of the equilibrium profiles due to its presence.
In brief, it increases outside the MI and decreases inside the MI following the pressure gradient. 
This implies that the turbulent transport around the MI can be also modified, meaning that the evolution of the MI and the evolution of plasma turbulence are coupled.
However, there has been relatively little experimental research progress~\cite{Bardoczi:2017gq, Chen:2017kb, Choi:2018uv, Jiang:2019fi} in addressing the effects of the modified plasma turbulence on the MI evolution to date.

Here, we report on experimental observations explaining how the modified turbulence can affect the nonlinear MI evolution. 
Firstly, it is found that the increase of electron temperature ($T_\mathrm{e}$) turbulence outside the MI can be limited and localized in the small region by the flow shear.
This could lead to the less turbulent transport into the MI from the inner (hotter) region, increasing the $J_\mathrm{BS}$ loss.
Combined with the dynamics of the flow shear development, it is shown that a positive feedback loop can exist for the NTM MI growth.
On the other hand, when the localized turbulence strength increases sufficiently to overcome the flow shear, the electron heat transport from the inner region into the MI is enhanced.
This leads to a beneficial $T_\mathrm{e}$ peaking inside the MI, which can decrease the MI width.
Finally, the further enhanced turbulence leads to a violent plasma disruption by expediting the magnetic reconnection and field line stochastization.
Compared to the case without the turbulence, a time scale of the turbulence-associated disruption is about one order smaller. 
These observations significantly extend our understanding of the nonlinear MI evolution in tokamak as well as provide general insights into the MI evolution physics in magnetized plasmas. 

\section*{Inhomogeneous low-$k$ $T_\mathrm{e}$ turbulence}

While turbulence around the MI has been limitedly observed in other experiments, the Korea Superconducting Tokamak Advanced Research (KSTAR)~\cite{Park:2019dv} experiment~\cite{Choi:2017ez} clearly demonstrates that the increase of the low-$k$ ($k\rho_\mathrm{i} < 1$ where $k$ is the wavenumber and $\rho_\mathrm{i}$ is the ion Larmor radius) $T_\mathrm{e}$ turbulence can be localized in the inner region ($r < r_\mathrm{s}$ where $r_\mathrm{s}$ is the MI boundary) between the X-point and O-point poloidal angles ($\theta_\mathrm{X} < \theta < \theta_\mathrm{O}$).
MIs in tokamak plasmas can grow either spontaneously through the TM~\cite{Furth:1963hd} or NTM~\cite{Carrera:1986cca} instabilities, or be driven externally via the forced magnetic reconnection~\cite{Hahm:1985cz}. 
In the KSTAR experiment~\cite{Choi:2017ez}, the $m/n=2/1$ MI was driven and held by the external $n=1$ magnetic field perturbation for an accurate measurement of small $T_\mathrm{e}$ fluctuations around the MI using the 2D local $T_\mathrm{e}$ diagnostics.
Here, $m$ and $n$ are the poloidal and toroidal mode number, respectively. 

\begin{figure}[t]
\includegraphics[keepaspectratio,width=0.45\textwidth]{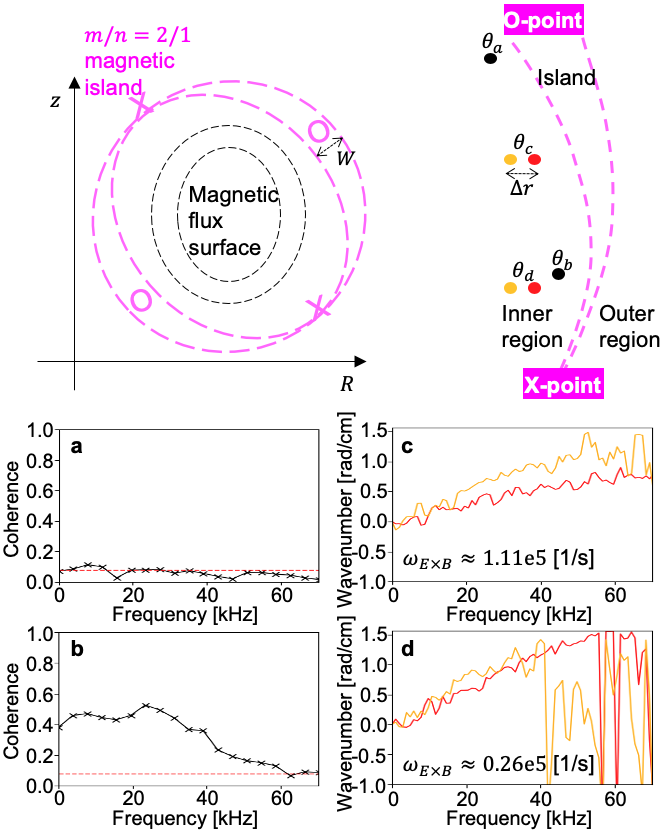}
\caption{\textbf{a}--\textbf{d}, The normalized frequency spectra (\textbf{a}, \textbf{b}) and local dispersion relations (\textbf{c}, \textbf{d}) of the low-$k$ $T_\mathrm{e}$ turbulence measured at $\theta_a$, $\theta_b$, $\theta_c$, and $\theta_d$, respectively. Orange and red lines in \textbf{c} or \textbf{d} represent the two-point measurements at orange and red points separated radially.}
\label{fig:inhomo}
\end{figure}

The result of 2D measurements of the $T_\mathrm{e}$ fluctuation spectrum and flow shear in the inner region of the MI is summarized in figure~\ref{fig:inhomo}.
Figures~\ref{fig:inhomo}a and \ref{fig:inhomo}b show the normalized spectra (coherence) of the $T_\mathrm{e}$ fluctuation at $\theta_a = \theta_O$ and $\theta_X < \theta_b < \theta_O$, respectively.
Fluctuation components whose power exceeding the significance level (red dashed line) are not observed at the O-point angle, while the strong broadband fluctuations are observed in the region a little distant from the X-point. 
It means that the $T_\mathrm{e}$ turbulence does not increase in the inner region at the O-point angle where the temperature profile steepened most, even though its amplitude depends on the $T_\mathrm{e}$ gradient (see figure~\ref{fig:spreading}a).
This can be understood as an effect of the inhomogeneous flow shear around the MI. 
The strong flow shearing rate can suppress the turbulence growth~\cite{Biglari:1990hxa, Hahm:1995eb}, and it was found that the flow shearing rate is strongly increasing towards the O-point angle in the inner region.
Figures~\ref{fig:inhomo}c and \ref{fig:inhomo}d show radial two-point measurements of the local average dispersion relation ($k_z(f)$) at $\theta_c$ and $\theta_d$, respectively, where $\theta_X < \theta_d < \theta_c < \theta_O$.
We can assume that the intrinsic phase velocity of turbulence is uniform in the small ($\Delta r \ll r$) measurement region.
Then, the radial variation of the laboratory frame phase velocity ($v_\mathrm{L} = 2 \pi f / k_z$) is attributed to the radial variation of the background $E \times B$ flow ($v_{E\times B}$). 
The radial shearing rate of the $E \times B$ flow ($\omega_{E\times B}$) can be obtained approximately using local two-point measurements of $v_\mathrm{L}$, i.e. $\omega_{E\times B} = \Delta v_{E \times B} / \Delta r \approx \Delta v_\mathrm{L} / \Delta r$.
The measurements show that $\omega_{E\times B}$ is strongly increasing towards the O-point angle ($\omega_{E\times B} \approx 0.26 \pm 0.15 \times 10^5$ [1/s] at $\theta_d$ and $\omega_{E\times B} \approx 1.11 \pm 0.48 \times 10^5$ [1/s] at $\theta_c$)~\cite{Choi:2017ez}.
This offers an explanation of the absence of the significant fluctuation in the inner region at the O-point angle since $\omega_{E\times B}$ is expected to be larger than the typical auto-decorrelation rate ($10^5$ [1/s]) of tokamak plasma turbulence.
Though it is not shown in this paper, the flow shearing rate across the MI is also large enough ($\omega_{E\times B} \ge 1.0 \times 10^5$ [1/s]) for preventing the increased turbulence near the X-point from spreading into the MI in this case~\cite{Choi:2017ez}.
Note that these spatial patterns of the turbulence and the shear flow are in broad agreements with the results from the fluid and gyrokinetic simulations~\cite{Poli:2009ce, Hornsby:2010fh, Izacard:2016de, Hu:2016kea, Kwon:2018bi, Fang:2019fz}.
In particular, reference~\cite{Kwon:2018bi} contains gyrokinetic simulation results validated against the experiments. 
In spite of some limitations, a main role of the increasing flow shear toward the O-point in suppressing the turbulence was confirmed.

\section*{A threshold behavior and positive feedback loop}

The strong flow shear, developing around the MI, would suppress turbulence and affect the turbulent transport across the MI. 
With the suppressed turbulence and the reduced heat influx from the inner (hotter) region, the resulting profile would be sharper across the MI boundary and flatter inside the MI than the profile with the stronger turbulence. 
The flatter profile inside the MI is expected to increase the $J_\mathrm{BS}$ loss and enhance the NTM growth rate.

\begin{figure}[t]
\includegraphics[keepaspectratio,width=0.45\textwidth]{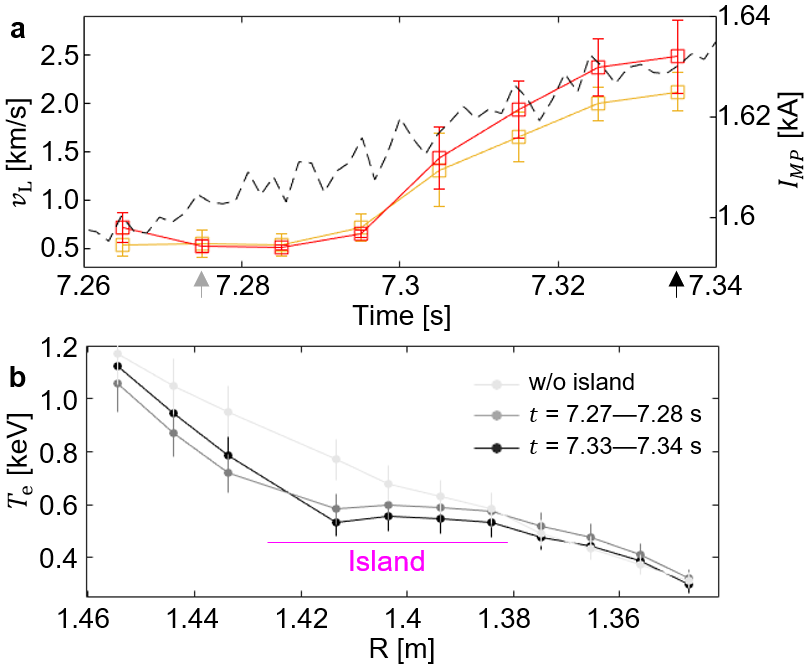}
\caption{Flow shear development. \textbf{a}, The laboratory phase velocity ($v_\mathrm{L}$) measurements in time at two radial positions in the inner region (red and orange squares) with the slowly increased external field (dashed black line). \textbf{b}, Mean $T_\mathrm{e}$ profiles measured at the high field side for different periods.}
\label{fig:flow}
\end{figure}

It is worth noting that the flow shear development by the MI has shown a threshold behavior.
In the experiment, as the external magnetic field perturbation is slowly increasing, the flow shear (or the difference between local two-point measurements of $v_\mathrm{L}$) starts to develop above a certain value of the external field as shown in figure~\ref{fig:flow}a.
It also seems to increase with the external field beyond the threshold value.
Note that error bars of $v_\mathrm{L}$ represent the standard deviation of measurements for many $k_z$s in the broadband local dispersion relation. 
Though the resolution of $T_\mathrm{e}$ profile measurements shown in figure~\ref{fig:flow}b is not sufficient to detect the change of the MI width, the MI width is expected to slightly increase with the strength of the perturbation.
The flow shear development with a large magnetic island and its increasing trend with the island width are also observed in other experimental~\cite{Choi:2017ez, Ida:2002ga} and simulation~\cite{Navarro:2017ei} studies.

\begin{figure}[t]
\includegraphics[keepaspectratio,width=0.45\textwidth]{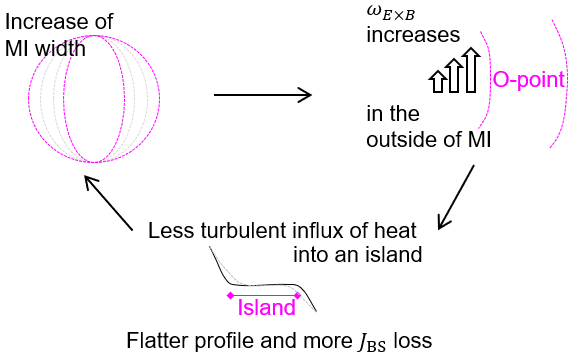}
\caption{A positive feedback loop for the NTM growth.}
\label{fig:loop}
\end{figure}

Comparing $T_\mathrm{e}$ profiles before and after the flow shear development, a sharper $T_\mathrm{e}$ profile across the inner boundary is found with a lower $T_\mathrm{e}$ value inside the MI, which can be the result of the reduced heat transport from the inner region with the increased flow shear. 
In the low-$\beta_\theta$ KSTAR experiment, the MI was driven by the external field and the effect of the reduced transport on the MI evolution may not be critical.
However, for the NTM in the high-$\beta_\theta$ plasma, the reduced electron heat transport would play a significant role via the dominant $J_\mathrm{BS}$ loss ($\propto \beta_\theta$) term~\cite{Bardoczi:2017gq} since its effect can be amplified through a positive feedback loop as presented in figure~\ref{fig:loop}. 
As an MI width exceeds a threshold value, the flow shear develops across the MI and increases with the MI width. 
It quenches the turbulent influx of heat and particle into the MI, which results in the sharp boundary and a flatter pressure profile inside the MI. 
The $J_\mathrm{BS}$ loss is further enhanced, and the growth of the MI width is accelerated. 

The threshold behavior and proposed feedback loop suggested in the previous paragraph should be considered for a more complete understanding of the NTM growth.
There would be a width or a range of width in which the shearing rate ($\omega_\mathrm{E \times B}(W)$) of the flow exceeds the turbulence growth rate driven by the local pressure gradient ($\Delta \omega(W)$).
The MI with $\omega_\mathrm{E \times B}(W) > \Delta \omega(W)$ can grow further via the suggested feedback loop.
In the recent DIII-D~\cite{Buttery:2018gr} experiment~\cite{Choi:2018uv}, the density turbulence modification due to the NTM MI is observed in the particular period of the MI growing phase. 
Previous understanding on the modulation of density turbulence around the magnetic island relied on the flattening of the pressure profile inside the large island~\cite{Bardoczi:2016gj}.
However, this observation is limited for the particular range of the island width in the growing phase.
The suggested feedback loop associated with $\omega_\mathrm{E \times B}(W) > \Delta \omega(W)$ condition for turbulence suppression provides a reasonable explanation for this observation.
Note that, however, different patterns of the turbulence and the flow were also observed in other experiments, and even a transition between a strong and weak $\omega_\mathrm{E \times B}$ states (low and high accessibility states~\cite{Ida:2015fm}, respectively) is observed. 

\section*{Stabilizing effect of the low-$k$ turbulence via turbulence spreading}

While turbulence can be suppressed by the increased flow shear near the O-point angle, it remains significant near the X-point due to the finite $T_\mathrm{e}$ gradient and the insufficient flow shear there.
The increased turbulence would be localized in the inner region when its amplitude is too small to overcome the flow shear across the MI~\cite{Choi:2017ez}. 
In other words, this turbulence outside the MI can spread into the MI if its amplitude is sufficiently large~\cite{Wang:2007jg}.
In the recent DIII-D~\cite{Ida:2018kx} and HL-2A~\cite{Jiang:2019fi} experiments, the spreading of the density and temperature turbulence into the O-point of the MI are observed, respectively.   

Turbulence spreading~\cite{Garbet:1994fx, Hahm:2004kb, Grenfell:2020iy} is expected to play an important role for the MI evolution~\cite{Hahm:2018dm}, since the accompanying heat or particle flux can change the pressure profile inside the MI.
This can result in the smaller MI saturation width and the stabilization of the NTM via the Ohmic current perturbation~\cite{Hegna:1998fi, Classen:2007kj} and the recovery of the pressure gradient (and $J_\mathrm{BS}$)~\cite{Bardoczi:2017gq}, respectively.
Detailed observation of the dynamics of turbulence spreading would be helpful to understand its effect on the MI evolution.
\begin{figure}
\includegraphics[keepaspectratio,width=0.49\textwidth]{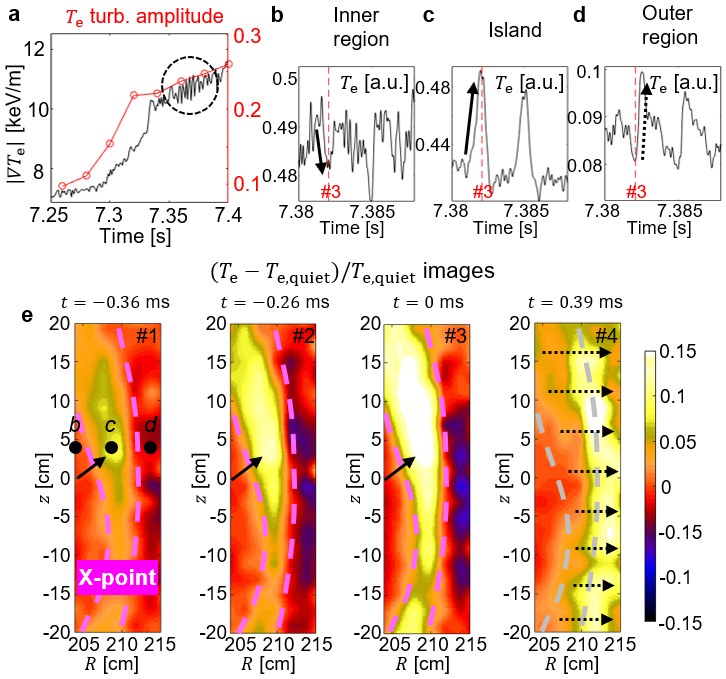}
\caption{Turbulence spreading and temperature peaking. \textbf{a}, The $T_\mathrm{e}$ gradient and low-$k$ $T_\mathrm{e}$ turbulence amplitude near \textit{b} (marked on \#1 in \textbf{e}) in the inner region with the increased external field. Intermittent turbulence spreading events are observed during the period indicated by a dashed circle. \textbf{b}--\textbf{d}, $T_\mathrm{e}$ time traces at \textit{b}, \textit{c}, and \textit{d}, respectively. \textbf{e}, The 2D normalized change of $T_\mathrm{e}$ around the MI during a single event. Absolute time of \#3 is indicated by a red dashed line in \textbf{b}--\textbf{d}.}
\label{fig:spreading}
\end{figure}
An experimental observation, which can be interpreted as an intermittent turbulence spreading event, is introduced below with detailed measurements of its dynamics.
The nonlinear electrostatic gyrokinetic simulation based on the KSTAR plasma equilibrium and profiles showed that the turbulence and heat can spread into the MI which was initially stable region~\cite{Kwon:2018ui}.

Figure~\ref{fig:spreading}a shows evolution of the $T_\mathrm{e}$ gradient and low-$k$ turbulence amplitude in the inner region of the MI near the X-point as the external magnetic field perturbation increases. 
Both the gradient and the turbulence amplitude increase in time following the external field.
Initially, the increased turbulence is localized in the outside of the MI due to the strong flow shear across the MI~\cite{Choi:2017ez}.
When the turbulence amplitude is larger than a certain critical value, intermittent heat transport events are observed repeatedly (indicated by a dashed circle in figure~\ref{fig:spreading}a). 
The time traces of local $T_\mathrm{e}$ at \textit{b} in the inner region, at \textit{c} inside the MI, and at \textit{d} in the outer region are also shown in figures~\ref{fig:spreading}b, \ref{fig:spreading}c, and \ref{fig:spreading}d, respectively. 
Bold arrows in figures~\ref{fig:spreading}b and \ref{fig:spreading}c indicate the enhanced heat transport from the inner region (where $T_\mathrm{e}$ decreases) into the MI (where $T_\mathrm{e}$ increases) during one event, and a dashed arrow in figure~\ref{fig:spreading}d indicates the rapid and global exhaust of the heat accumulated inside the MI.

\begin{figure}
\includegraphics[keepaspectratio,width=0.49\textwidth]{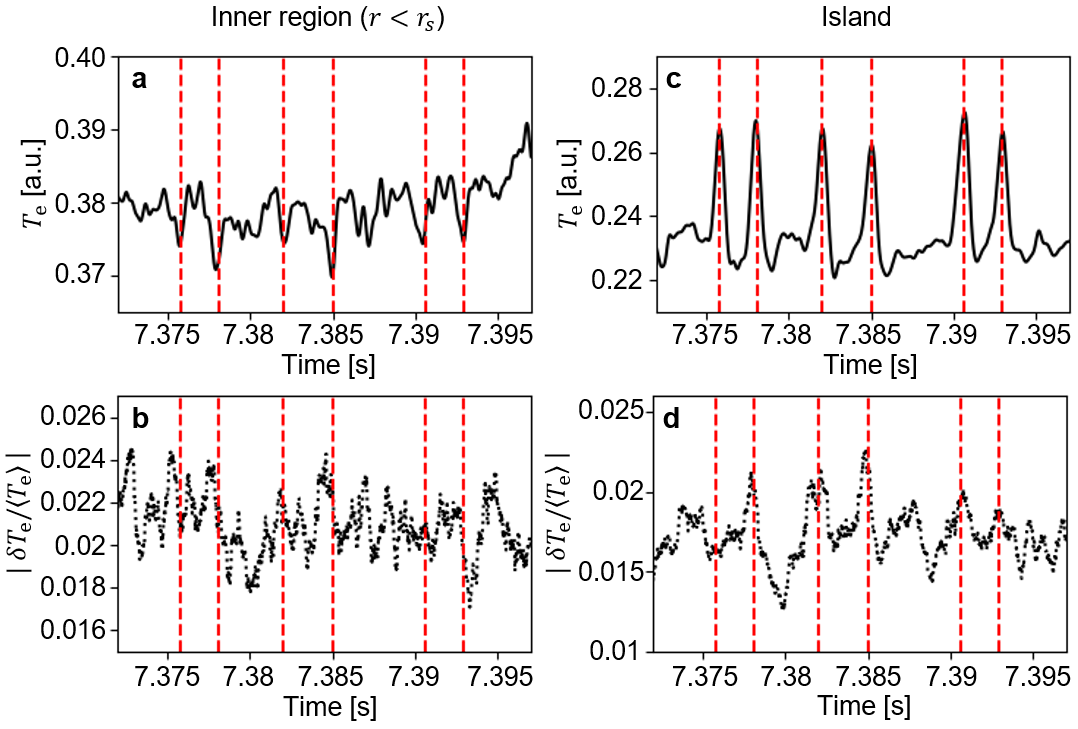}
\caption{\textbf{a}--\textbf{d}, A low-pass filtered ($f<1$~kHz) $T_\mathrm{e}$ signal (\textbf{a}, \textbf{c}) and the RMS amplitude of a band-pass filtered ($30 \le f \le 70$~kHz) $\delta T_\mathrm{e}/\langle T_\mathrm{e} \rangle$ (\textbf{b}, \textbf{d}) outside and inside the magnetic island, respectively.}
\label{fig:tbspread}
\end{figure}

Four images (\#1--\#4) in figure~\ref{fig:spreading}e show the local change of the 2D $T_\mathrm{e}$ during a single event. 
From \#1 to \#3, $T_\mathrm{e}$ increases inside the MI spontaneously. 
This increase can be a result of the intermittent leakage of the turbulent heat flux from the inner region through a path near the X-point (weak flow shear).
In other words, the heat transport can be enhanced by turbulence spreading, which can carry the heat flux into the MI and result in a modest peaking of $T_\mathrm{e}$ inside the MI.
Note that the interior magnetic topology of the MI is known to have a good perpendicular confinement characteristics~\cite{Inagaki:2004dn} to sustain the peaking.
Figure~\ref{fig:tbspread} shows how the low-$k$ turbulence amplitude varies during the event in higher temporal resolution.
Most events are found to be correlated with the RMS amplitude ($|\delta T_\mathrm{e} / \langle T_\mathrm{e} \rangle |$) of the low-$k$ broadband (30--70~kHz) turbulence.
$|\delta T_\mathrm{e} / \langle T_\mathrm{e} \rangle |$ in the inner region (shown in figure~\ref{fig:tbspread}b) seems to start to decrease with the event and $|\delta T_\mathrm{e} / \langle T_\mathrm{e} \rangle |$ inside the island (shown in figure~\ref{fig:tbspread}d) is peaked with the event, which is consistent with a picture of turbulence spreading. 
The former is relatively clear in the 1st, 2nd, 4th, and 6th event, and the later is in the 2nd, 3rd, 4th, 5th, and 6th event.
The rapid and global exhaust of the accumulated heat inside the MI, which is observed for \#3--\#4, can be attributed to the global enhancement of the heat transport with the increased turbulence inside the MI. 

Although this observation is made with the externally driven MI, what is shown in figure~\ref{fig:spreading} would represent a general feature of turbulence spreading dynamics across the MI.
It implies two potentially beneficial effects of turbulence spreading on TM/NTM stability.
Firstly, the $T_\mathrm{e}$ peaking inside the MI can be beneficial for the MI saturation through its perturbation on the Ohmic current~\cite{Hegna:1998fi} as demonstrated in the auxiliary heating experiment~\cite{Classen:2007kj}.
In case of the MI driven by the NTM, the enhanced turbulent transport into the MI would eventually lead to a saturation at the smaller width or a stabilization, since it can recover the $J_\mathrm{BS}$ loss~\cite{Bardoczi:2017gq}.
Indeed, a partial stabilization of the NTM with the pellet injection was observed in the recent DIII-D and KSTAR experiments~\cite{Bardoczi:2019jw} in which the spreading of density turbulence~\cite{Ida:2018kx} was regarded as a key mechanism for the stabilization. 

\section*{Roles of the low-$k$ turbulence in the fast disruption}

Finally, we report an observation during the disruption phase of the MI, demonstrating a crucial role of the increased turbulence in the magnetic reconnection physics.
MIs in tokamak plasmas often lead to the disruption when it grows large enough. 
There are different kinds of the MI-associated disruption characterized by different temporal and spatial scales~\cite{Hender:2007ki}.
The explosive and massive disruption involving the fast magnetic reconnection is the most dangerous one.  
Interestingly, a violent minor disruption is observed along with the significantly enhanced turbulence at the X-point. 

\begin{figure}
\includegraphics[keepaspectratio,width=0.45\textwidth]{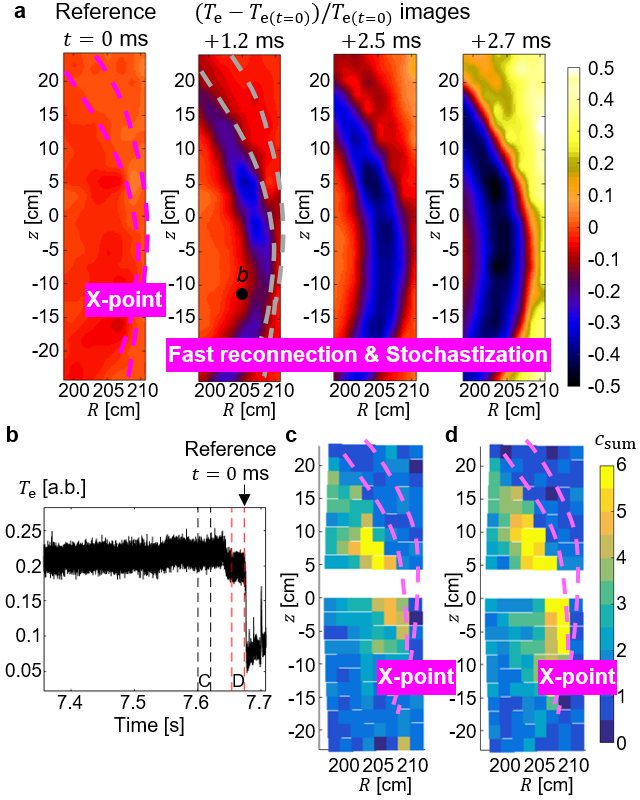}
\caption{Evolution of turbulence during disruption. \textbf{a}, The 2D normalized change of $T_\mathrm{e}$ around the MI during the fast disruption. \textbf{b}, $T_\mathrm{e}$ time trace at \textit{b} (marked on the second image in \textbf{a}). \textbf{c},\textbf{d}, The 2D strength of the low-$k$ $T_\mathrm{e}$ turbulence for the period C and D, respectively.}
\label{fig:disruption}
\end{figure}
Figure~\ref{fig:disruption}a shows the process of the fast MI disruption captured by the 2D $T_\mathrm{e}$ diagnostics, and time trace of $T_\mathrm{e}$ measured at \textit{b} in the inner region is shown in figure~\ref{fig:disruption}b. 
The constant external field is used to induce the MI in this experiment. 
The 2D $T_\mathrm{e}$ images represent the local change of $T_\mathrm{e}$ against the reference time ($t=0$) which is just before the disruption. 
The image at $t=1.2$~ms shows a local $T_\mathrm{e}$ collapse (indicated by a blue color) in the inner region with a structure reminiscent of the original MI (whose boundaries are indicated by purple dashed lines in the reference image), indicating an inward expansion of the MI width.
The following images show that $T_\mathrm{e}$ collapses more globally, probably due to the global field line stochastization and reconnection.
Figures~\ref{fig:disruption}c and \ref{fig:disruption}d are the images of the $T_\mathrm{e}$ coherence sum ($c_\mathrm{sum}=\sum_f c(f)$) for periods of C (quasi-steady state) and D (just before the disruption), respectively. 
They represent the local strength of the low-$k$ broadband $T_\mathrm{e}$ turbulence in each period.
In the period C, the turbulence is localized in a distant region from the X-point as shown in figure~\ref{fig:disruption}c. 
In the period D, the turbulence strength increases and it expands poloidally and reaches to the X-point as shown in figure~\ref{fig:disruption}d. 
It is noteworthy that this low-$k$ $T_\mathrm{e}$ turbulence is only observed for the fast disruption whose time scale is a few milliseconds.
Other similar MI-associated disruptions occur in a longer timescale (5--10 times) without notable low-$k$ turbulence~\cite{Kim:2018uy}. 

The proximity of the increased turbulence to the X-point during the fast disruption indicates a role of turbulence in the disruption processes, i.e. the magnetic reconnection and the field line stochastization.
It is generally known that anomalous dissipations by turbulence can enhance the magnetic reconnection~\cite{Itoh:1999kt} and realignment of global magnetic fields. 
On the other hand, a fast disruption driven by successive formation and coalescence of plasmoid-like structures~\cite{Wang:2016ji} was also observed in the KSTAR experiment~\cite{Choi:2016jz}.
However, the low-$k$ fluctuations in $T_\mathrm{e}$ discussed in this paper was not observed in the plasmoid-like-structure-associated disruption. 
The plasmoid-like structure might be attributed to the density turbulence. 

In summary, this paper reports on experimental observations which indicate that the modified turbulence around an MI can either accelerate the MI growth via the reduced influx of heat into the MI, or lead to a saturation or stabilization of the MI via turbulence spreading in tokamak plasmas. 
Turbulence can also lead to a violent plasma disruption by expediting the magnetic reconnection and forming the stochastic region at the X-point.

\section*{Acknowledgements}

The authors would appreciate all the supports from the KSTAR team.
This research was supported by Korean Ministry of Science and ICT under NFRI R\&D programs (NFRI-EN1901-10 and NFRI-EN1941-5), and also by National Research Foundation of Korea under NRF-2019M1A7A1A03088462.

\section*{Author contributions}

M.J.C. developed the central idea of the research based on discussion with L.B., J.-M.K., T.S.H., H.K.P. and E.Y. about the previous experiment conducted by J.K.. 
M.J.C. gathered the data and carried out the further experiment with general guidance from J.K., M.W. and B.-H.P.. 
L.B., J.-M.K., T.S.H., H.K.P. and E.Y. contributed to the interpretation of the experimental results and the revision of the draft.
G.S.Y. contributed to the acquisition of the diagnostics data presented in this article.

\section*{Competing interests}

The authors declare no competing interests.

\section*{METHODS}

\subsection*{The externally driven magnetic island}

In tokamak plasmas, MIs can be driven at the rational $q$ surface by the external magnetic field perturbation which has a resonant component to that rational surface.
The externally driven MI is locked in the position by the external field, which allows an accurate measurement of dynamics of the MI and ambient turbulence.
For the externally driven MI experiment on KSTAR, the $n=1$ external magnetic field perturbation was used to drive the $m/n=2/1$ tearing mode MI at the $q=2$ flux surface.
The KSTAR plasma of the driven MI experiment has the major radius $R=180$~cm, the minor radius $a\sim 50$~cm, the toroidal field $B_\mathrm{T}=2.0$--2.2~T, the plasma current $I_\mathrm{p}=600$--700~kA, the Spitzer resistivity $\eta \sim 1.2\times10^{-7}$~Ohm$\cdot$m, and $\beta_\theta \sim 0.5$~\%.

\subsection*{The 2D $T_\mathrm{e}$ diagnostics}

Tokamak core plasma can be optically thick for the fundamental O-mode or the second harmonic X-mode of the electron cyclotron emission (ECE). 
The measured intensity of the optically thick ECE depends linearly on the local electron temperature by Kirchhoff's law and Rayleigh-Jeans law. 
On the other hand, the ECE frequency in tokamak plasma follows $1/R$ dependence of the toroidal field.
A heterodyne detector can measure the radial profile of the electron temperature by measuring the ECE intensity selectively in frequency space. 
The electron cyclotron emission imaging (ECEI) diagnostics was developed to measure the local 2D $T_\mathrm{e}$ in ($R,z$) space using a vertical array of heterodyne detectors. 
The ECEI diagnostics on the KSTAR tokamak has 24 heterodyne detectors in vertical direction and each detector has 8 radial channels, i.e. total 192 channels~\cite{Yun:2014kv}. 
It can measure the local 2D $T_\mathrm{e}$ with a high spatial ($\Delta R \approx \Delta z \le 2$~cm) and temporal ($\Delta t =0.5$--2~$\mu$s) resolution.
This diagnostics has been utilized to study various tokamak plasma phenomena from the magnetohydrodynamic instabilities to the low-$k$ broadband turbulence~\cite{Park:2019kq}.
For example, measurements of the local 2D electron temperature have revealed the change of the magnetic field topology successfully in various magnetic reconnection events of magnetohydrodynamic instabilities~\cite{Choi:2016jz, Park:2006il}. 

\section*{Data availability}

Raw data were generated at the KSTAR facility. 
Derived data are available from the corresponding author upon request.

\section*{Code availability}

The data analysis codes used for figures of this article are available via the GitHub repository~\url{https://github.com/minjunJchoi/fluctana}~\cite{Choi:2019tw}.

\bibliographystyle{ieeetr}

\begin{thebibliography}{10}

\bibitem{Furth:1963hd}
H.~P. Furth, J.~Killeen, and M.~N. Rosenbluth, ``{Finite-Resistivity
  Instabilities of a Sheet Pinch},'' {\em Physics of Fluids}, vol.~6, no.~4,
  pp.~459--459, 1963.

\bibitem{Carrera:1986cca}
R.~Carrera, R.~D. Hazeltine, and M.~Kotschenreuther, ``{Island bootstrap
  current modification of the nonlinear dynamics of the tearing mode},'' {\em
  Physics of Fluids}, vol.~29, no.~4, p.~899, 1986.

\bibitem{Chang:1995una}
Z.~Chang, J.~D. Callen, E.~D. Fredrickson, R.~V. Budny, C.~C. Hegna, K.~M.
  McGuire, and M.~C. Zarnstorff, ``{Observation of nonlinear neoclassical
  pressure-gradient-driven tearing modes in TFTR},'' {\em Physical Review
  Letters}, vol.~74, no.~23, p.~4663, 1995.

\bibitem{Haye:2006gh}
R.~J. La~Haye, ``{Neoclassical tearing modes and their control},'' {\em Physics
  of Plasmas}, vol.~13, no.~5, p.~055501, 2006.

\bibitem{Zhao:2015gra}
K.~J. Zhao, Y.~J. Shi, S.~H. Hahn, P.~H. Diamond, Y.~Sun, J.~Cheng, H.~Liu,
  N.~Lie, Z.~P. Chen, Y.~H. Ding, Z.~Y. Chen, B.~Rao, M.~Leconte, J.~G. Bak,
  Z.~F. Cheng, L.~Gao, X.~Q. Zhang, Z.~J. Yang, N.~C. Wang, L.~Wang, W.~Jin,
  L.~W. Yan, J.~Q. Dong, G.~Zhuang, and {J-TEXT team}, ``{Plasma flows and
  fluctuations with magnetic islands in the edge plasmas of J-TEXT tokamak},''
  {\em Nuclear Fusion}, vol.~55, no.~7, p.~073022, 2015.

\bibitem{Rea:2015he}
C.~Rea, N.~Vianello, M.~Agostini, R.~Cavazzana, G.~De~Masi, E.~Martines,
  B.~Momo, P.~Scarin, S.~Spagnolo, G.~Spizzo, M.~Spolaore, and M.~Zuin,
  ``{Comparative studies of electrostatic turbulence induced transport in
  presence of resonant magnetic perturbations in RFX-mod},'' {\em Nuclear
  Fusion}, vol.~55, no.~11, p.~113021, 2015.

\bibitem{Estrada:2016gz}
T.~Estrada, E.~Ascas{\'\i}bar, E.~Blanco, A.~Cappa, C.~Hidalgo, K.~Ida,
  A.~L{\'o}pez-Fraguas, and B.~P. van Milligen, ``{Plasma flow, turbulence and
  magnetic islands in TJ-II},'' {\em Nuclear Fusion}, vol.~56, no.~2,
  p.~026011, 2016.

\bibitem{Bardoczi:2016gj}
L.~Bard{\'o}czi, T.~L. Rhodes, T.~A. Carter, A.~Ba{\~n}{\'o}n~Navarro, W.~A.
  Peebles, F.~Jenko, and G.~McKee, ``{Modulation of Core Turbulent Density
  Fluctuations by Large-Scale Neoclassical Tearing Mode Islands in the DIII-D
  Tokamak},'' {\em Physical Review Letters}, vol.~116, no.~21, p.~215001, 2016.

\bibitem{Bardoczi:2017gq}
L.~Bard{\'o}czi, T.~A. Carter, R.~J. La~Haye, T.~L. Rhodes, and G.~R. McKee,
  ``{Impact of neoclassical tearing mode{\textendash}turbulence multi-scale
  interaction in global confinement degradation and magnetic island
  stability},'' {\em Physics of Plasmas}, vol.~24, no.~12, p.~122503, 2017.

\bibitem{Choi:2017ez}
M.~J. Choi, J.~Kim, J.~M. Kwon, H.~K. Park, Y.~In, W.~Lee, K.~D. Lee, G.~S.
  Yun, J.~Lee, M.~Kim, W.~H. Ko, J.~H. Lee, Y.~S. Park, Y.~S. Na, N.~C.
  Luhmann~Jr, and B.~H. Park, ``{Multiscale interaction between a large scale
  magnetic island and small scale turbulence},'' {\em Nuclear Fusion}, vol.~57,
  no.~12, p.~126058, 2017.

\bibitem{Chen:2017kb}
W.~Chen, M.~Jiang, Y.~Xu, P.~W. Shi, L.~M. Yu, X.~T. Ding, Z.~B. Shi, X.~Q. Ji,
  D.~L. Yu, Y.~G. Li, Z.~C. Yang, W.~L. Zhong, Z.~Y. Qiu, J.~Q. Li, J.~Q. Dong,
  Q.~W. Yang, Y.~Liu, L.~W. Yan, M.~Xu, and X.~R. Duan, ``{Experimental
  observation of multi-scale interactions among kink/tearing modes and
  high-frequency fluctuations in the HL-2A core NBI plasmas},'' {\em Nuclear
  Fusion}, vol.~57, no.~11, p.~114003, 2017.

\bibitem{Jiang:2018fz}
M.~Jiang, W.~L. Zhong, Y.~Xu, Z.~B. Shi, W.~Chen, X.~Q. Ji, X.~T. Ding, Z.~C.
  Yang, P.~W. Shi, A.~S. Liang, J.~Wen, J.~Q. Li, Y.~Zhou, Y.~G. Li, D.~L. Yu,
  Y.~Liu, Q.~W. Yang, and t.~H.-A. Team, ``{Influence of m/n=2/1 magnetic
  islands on perpendicular flows and turbulence in HL-2A Ohmic plasmas},'' {\em
  Nuclear Fusion}, vol.~58, no.~2, p.~026002, 2018.

\bibitem{Choi:2018uv}
M.~J. Choi, J.-M. Kwon, S.~H. Ko, J.~Kim, M.~H. Woo, K.~D. Lee, H.~K. Park,
  Y.~In, T.~S. Hahm, and G.~S. Yun, ``{Effect of multiscale interaction between
  an m/n=2/1 mode and micro instabilities on transport of KSTAR plasmas},'' in
  {\em The 27th IAEA Fusion Energy Conference}, pp.~EX/11--2, 2018.

\bibitem{Jiang:2019fi}
M.~Jiang, Y.~Xu, W.~Chen, Z.~Shi, J.~Li, W.~Wang, Z.~H. Qin, X.~Ding, W.~Zhong,
  X.~Q. Ji, P.~Shi, Z.~Yang, B.~Yuan, Y.~Liu, Q.~Yang, and M.~Xu, ``{Localized
  modulation of turbulence by m/n=1/1 magnetic islands in the HL-2A tokamak},''
  {\em Nuclear Fusion}, vol.~59, p.~066019, 2019.

\bibitem{Sun:2018hh}
P.~J. Sun, Y.~D. Li, Y.~Ren, X.~D. Zhang, G.~J. Wu, B.~Lyu, T.~H. Shi, L.~Q.
  Xu, F.~D. Wang, Q.~Li, J.~Z. Zhang, L.~Q. Hu, J.~G. Li, and t.~E. team,
  ``{Experimental study of the effect of 2/1 classical tearing mode on
  (intermediate, small)-scale microturbulence in the core of an EAST L mode
  plasma},'' {\em Plasma Physics and Controlled Fusion}, vol.~60, no.~2,
  p.~025019, 2018.

\bibitem{Connor:2009je}
H.~R. Wilson and J.~W. Connor, ``{The influence of magnetic islands on drift
  mode stability in magnetized plasma},'' {\em Plasma Physics and Controlled
  Fusion}, vol.~51, no.~11, p.~115007, 2009.

\bibitem{Poli:2009ce}
E.~Poli, A.~Bottino, and A.~G. Peeters, ``{Behaviour of turbulent transport in
  the vicinity of a magnetic island},'' {\em Nuclear Fusion}, vol.~49, no.~7,
  p.~075010, 2009.

\bibitem{Hornsby:2010fh}
W.~A. Hornsby, A.~G. Peeters, A.~P. Snodin, F.~J. Casson, Y.~Camenen,
  G.~Szepesi, M.~Siccinio, and E.~Poli, ``{The nonlinear coupling between
  gyroradius scale turbulence and mesoscale magnetic islands in fusion
  plasmas},'' {\em Physics of Plasmas}, vol.~17, no.~9, p.~092301, 2010.

\bibitem{Ichiguchi:2015hr}
K.~Ichiguchi, Y.~Suzuki, M.~Sato, Y.~Todo, T.~Nicolas, S.~Sakakibara,
  S.~Ohdachi, Y.~Narushima, and B.~A. Carreras, ``{Three-dimensional MHD
  analysis of heliotron plasma with RMP},'' {\em Nuclear Fusion}, vol.~55,
  no.~7, p.~073023, 2015.

\bibitem{Zarzoso:2015fh}
D.~Zarzoso, W.~A. Hornsby, E.~Poli, F.~J. Casson, A.~G. Peeters, and S.~Nasr,
  ``{Impact of rotating magnetic islands on density profile flattening and
  turbulent transport},'' {\em Nuclear Fusion}, vol.~55, no.~11, p.~113018,
  2015.

\bibitem{Izacard:2016de}
O.~Izacard, C.~Holland, S.~D. James, and D.~P. Brennan, ``{Dynamics of ion
  temperature gradient turbulence and transport with a static magnetic
  island},'' {\em Physics of Plasmas}, vol.~23, no.~2, p.~022304, 2016.

\bibitem{Hu:2016kea}
Z.~Q. Hu, Z.~X. Wang, L.~Wei, J.~Q. Li, and Y.~Kishimoto, ``{Dual roles of
  shear flow in nonlinear multi-scale interactions},'' {\em Nuclear Fusion},
  vol.~56, p.~016012, 2016.

\bibitem{Navarro:2017ei}
A.~Ba{\~n}{\'o}n~Navarro, L.~Bard{\'o}czi, T.~A. Carter, F.~Jenko, and T.~L.
  Rhodes, ``{Effect of magnetic islands on profiles, flows, turbulence and
  transport in nonlinear gyrokinetic simulations},'' {\em Plasma Physics and
  Controlled Fusion}, vol.~59, no.~3, pp.~034004--12, 2017.

\bibitem{Kwon:2018bi}
J.-M. Kwon, S.~Ku, M.~J. Choi, C.~S. Chang, R.~Hager, E.~S. Yoon, H.~H. Lee,
  and H.~S. Kim, ``{Gyrokinetic simulation study of magnetic island effects on
  neoclassical physics and micro-instabilities in a realistic KSTAR plasma},''
  {\em Physics of Plasmas}, vol.~25, no.~5, p.~052506, 2018.

\bibitem{Fang:2019fz}
K.~S. Fang and Z.~Lin, ``{Global gyrokinetic simulation of microturbulence with
  kinetic electrons in the presence of magnetic island in tokamak},'' {\em
  Physics of Plasmas}, vol.~26, no.~5, p.~052510, 2019.

\bibitem{Ishizawa:2019ky}
A.~Ishizawa, Y.~Kishimoto, and Y.~Nakamura, ``{Multi-scale interactions between
  turbulence and magnetic islands and parity mixture - a review},'' {\em Plasma
  Physics and Controlled Fusion}, vol.~61, p.~054006, 2019.

\bibitem{Yamada:2008fx}
T.~Yamada, S.-I. Itoh, T.~Maruta, N.~Kasuya, Y.~Nagashima, S.~Shinohara,
  K.~Terasaka, M.~Yagi, S.~Inagaki, Y.~Kawai, A.~Fujisawa, and K.~Itoh,
  ``{Anatomy of plasma turbulence},'' {\em Nature Physics}, vol.~4, no.~9,
  pp.~721--725, 2008.

\bibitem{Park:2019dv}
H.~K. Park, M.~J. Choi, S.~H. Hong, Y.~In, Y.~M. Jeon, J.~S. Ko, W.~H. Ko,
  J.~G. Kwak, J.~M. Kwon, J.~Lee, J.~H. Lee, W.~Lee, Y.~B. Nam, Y.~K. Oh, B.~H.
  Park, J.~K. Park, Y.~S. Park, S.~J. Wang, M.~Yoo, S.~W. Yoon, J.~G. Bak,
  C.~S. Chang, W.~H. Choe, Y.~Chu, J.~Chung, N.~Eidietis, H.~S. Han, S.~H.
  Hahn, H.~G. Jhang, J.~W. Juhn, J.~H. Kim, K.~Kim, A.~Loarte, H.~H. Lee, K.~C.
  Lee, D.~Mueller, Y.~S. Na, Y.~U. Nam, G.~Y. Park, K.~R. Park, R.~A. Pitts,
  S.~A. Sabbagh, G.~S. Yun, and {the KSTAR team}, ``{Overview of KSTAR research
  progress and future plans toward ITER and K-DEMO},'' {\em Nuclear Fusion},
  vol.~59, no.~11, p.~112020, 2019.

\bibitem{Hahm:1985cz}
T.~S. Hahm and R.~M. Kulsrud, ``{Forced magnetic reconnection},'' {\em Physics
  of Fluids}, vol.~28, no.~8, p.~2412, 1985.

\bibitem{Biglari:1990hxa}
H.~Biglari, P.~H. Diamond, and P.~W. Terry, ``{Influence of sheared poloidal
  rotation on edge turbulence},'' {\em Physics of Fluids B: Plasma Physics},
  vol.~2, no.~1, pp.~1--4, 1990.

\bibitem{Hahm:1995eb}
T.~S. Hahm and K.~H. Burrell, ``{Flow shear induced fluctuation suppression in
  finite aspect ratio shaped tokamak plasma},'' {\em Physics of Plasmas},
  vol.~2, no.~5, pp.~1648--1651, 1995.

\bibitem{Ida:2002ga}
K.~Ida, N.~Ohyabu, T.~Morisaki, Y.~Nagayama, S.~Inagaki, K.~Itoh, Y.~Liang,
  K.~Narihara, A.~Y. Kostrioukov, B.~J. Peterson, K.~Tanaka, T.~Tokuzawa,
  K.~Kawahata, H.~Suzuki, A.~Komori, and {LHD Experimental Group},
  ``{Observation of plasma flow at the magnetic island in the large helical
  device},'' {\em Physical Review Letters}, vol.~88, no.~1, p.~015002, 2002.

\bibitem{Buttery:2018gr}
R.~J. Buttery, B.~Covele, J.~Ferron, A.~Garofalo, C.~T. Holcomb, T.~Leonard,
  J.~M. Park, T.~Petrie, C.~Petty, G.~Staebler, E.~J. Strait, and
  M.~Van~Zeeland, ``{DIII-D Research to Prepare for Steady State Advanced
  Tokamak Power Plants},'' {\em Journal of Fusion Energy}, vol.~38, no.~1,
  p.~72, 2018.

\bibitem{Ida:2015fm}
K.~Ida, T.~Kobayashi, T.~E. Evans, S.~Inagaki, M.~E. Austin, M.~W. Shafer,
  S.~Ohdachi, Y.~Suzuki, S.~I. Itoh, and K.~Itoh, ``{Self-regulated oscillation
  of transport and topology of magnetic islands in toroidal plasmas},'' {\em
  Scientific Reports}, vol.~5, p.~16165, 2015.

\bibitem{Wang:2007jg}
W.~X. Wang, T.~S. Hahm, W.~W. Lee, G.~Rewoldt, J.~Manickam, and W.~M. Tang,
  ``{Nonlocal properties of gyrokinetic turbulence and the role of EXB flow
  shear},'' {\em Physics of Plasmas}, vol.~14, no.~7, p.~072306, 2007.

\bibitem{Ida:2018kx}
K.~Ida, T.~Kobayashi, M.~Ono, T.~E. Evans, G.~R. McKee, and M.~E. Austin,
  ``{Hysteresis Relation between Turbulence and Temperature Modulation during
  the Heat Pulse Propagation into a Magnetic Island in DIII-D},'' {\em Physical
  Review Letters}, vol.~120, no.~24, p.~245001, 2018.

\bibitem{Garbet:1994fx}
X.~Garbet, L.~Laurent, A.~Samain, and J.~Chinardet, ``{Radial propagation of
  turbulence in tokamaks},'' {\em Nuclear Fusion}, vol.~34, no.~7,
  pp.~963--974, 1994.

\bibitem{Hahm:2004kb}
T.~S. Hahm, P.~H. Diamond, Z.~Lin, K.~Itoh, and S.~I. Itoh, ``{Turbulence
  spreading into the linearly stable zone and transport scaling},'' {\em Plasma
  Physics and Controlled Fusion}, vol.~46, no.~5A, pp.~A323--A333, 2004.

\bibitem{Grenfell:2020iy}
G.~Grenfell, B.~P. van Milligen, U.~Losada, T.~Estrada, B.~Liu, C.~Silva,
  M.~Spolaore, C.~Hidalgo, and t.~T.-I. Team, ``{The impact of edge radial
  electric fields on edge{\textendash}scrape-off layer coupling in the TJ-II
  stellarator},'' {\em Nuclear Fusion}, vol.~60, p.~014001, Jan. 2020.

\bibitem{Hahm:2018dm}
T.~S. Hahm and P.~H. Diamond, ``{Mesoscopic Transport Events and the Breakdown
  of Fick{\textquoteright}s Law for Turbulent Fluxes },'' {\em Journal of the
  Korean Physical Society}, vol.~73, no.~6, pp.~747--792, 2018.

\bibitem{Hegna:1998fi}
C.~C. Hegna and J.~D. Callen, ``{On the stabilization of neoclassical
  magnetohydrodynamic tearing modes using localized current drive or
  heating},'' {\em Physics of Plasmas}, vol.~4, no.~8, pp.~2940--2946, 1998.

\bibitem{Classen:2007kj}
I.~G.~J. Classen, E.~Westerhof, C.~Domier, A.~J.~H. Donn{\'e}, R.~Jaspers,
  N.~Luhmann, H.~Park, M.~van~de Pol, G.~Spakman, and M.~Jakubowski, ``{Effect
  of Heating on the Suppression of Tearing Modes in Tokamaks},'' {\em Physical
  Review Letters}, vol.~98, no.~3, p.~035001, 2007.

\bibitem{Kwon:2018ui}
J.-M. Kwon, S.~Ku, C.~S. Chang, M.~J. Choi, R.~Hager, E.~S. Yoon, H.~H. Lee,
  and H.~S. Kim, ``{Gyrokinetic XGC1 simulation study of magnetic island
  effects on neoclassical and turbulence physics in a KSTAR plasma},'' in {\em
  The 27th IAEA Fusion Energy Conference}, pp.~TH--8, 2018.

\bibitem{Inagaki:2004dn}
S.~Inagaki, N.~Tamura, K.~Ida, Y.~Nagayama, K.~Kawahata, S.~Sudo, T.~Morisaki,
  K.~Tanaka, T.~Tokuzawa, and {LHD Experimental Group}, ``{Observation of
  reduced heat transport inside the magnetic island O point in the large
  helical device},'' {\em Physical Review Letters}, vol.~92, no.~5, p.~055002,
  2004.

\bibitem{Bardoczi:2019jw}
L.~Bard{\'o}czi, M.~J. Choi, A.~Ba{\~n}{\'o}n~Navarro, D.~Shiraki, R.~J.
  La~Haye, S.~H. Park, M.~Knolker, T.~E. Evans, G.~R. McKee, M.~Woo, B.~H.
  Park, and F.~Jenko, ``{Controlled neoclassical tearing mode (NTM) healing by
  fueling pellets and its impact on electron cyclotron current drive
  requirements for complete NTM stabilization},'' {\em Nuclear Fusion},
  vol.~59, p.~126047, Dec. 2019.

\bibitem{Hender:2007ki}
T.~C. Hender, J.~C. Wesley, J.~Bialek, A.~Bondeson, A.~H. Boozer, R.~J.
  Buttery, A.~Garofalo, T.~P. Goodman, R.~S. Granetz, Y.~Gribov, O.~Gruber,
  M.~Gryaznevich, G.~Giruzzi, S.~G{\"u}nter, N.~Hayashi, P.~Helander, C.~C.
  Hegna, D.~F. Howell, D.~A. Humphreys, G.~T.~A. Huysmans, A.~W. Hyatt,
  A.~Isayama, S.~C. Jardin, Y.~Kawano, A.~Kellman, C.~Kessel, H.~R. Koslowski,
  R.~J.~L. Haye, E.~Lazzaro, Y.~Q. Liu, V.~Lukash, J.~Manickam, S.~Medvedev,
  V.~Mertens, S.~V. Mirnov, Y.~Nakamura, G.~Navratil, M.~Okabayashi, T.~Ozeki,
  R.~Paccagnella, G.~Pautasso, F.~Porcelli, V.~D. Pustovitov, V.~Riccardo,
  M.~Sato, O.~Sauter, M.~J. Schaffer, M.~Shimada, P.~Sonato, E.~J. Strait,
  M.~Sugihara, M.~Takechi, A.~D. Turnbull, E.~Westerhof, D.~G. Whyte,
  R.~Yoshino, and H.~Zohm, ``{MHD stability, operational limits and
  disruptions},'' {\em Nuclear Fusion}, vol.~47, no.~6, p.~S128, 2007.

\bibitem{Kim:2018uy}
J.~Kim, M.~J. Choi, A.~Y. Aydemir, Y.~In, H.~Han, J.~K. Park, J.~Lee, J.~G.
  Bak, W.~H. Ko, and Y.~K. Oh, ``{EX/P7-14 Evolution of locked mode in the
  presence of non-axisymmetric fields in KSTAR},'' in {\em The 27th IAEA Fusion
  Energy Conference}, pp.~EX/P7--14, 2018.

\bibitem{Itoh:1999kt}
S.~I. Itoh, K.~Itoh, H.~Zushi, and A.~Fukuyama, ``{Physics of collapse events
  in toroidal plasmas},'' {\em Plasma Physics and Controlled Fusion}, vol.~40,
  no.~6, pp.~879--929, 1999.

\bibitem{Wang:2016ji}
R.~Wang, Q.~Lu, R.~Nakamura, C.~Huang, A.~Du, F.~Guo, W.~Teh, M.~Wu, S.~Lu, and
  S.~Wang, ``{Coalescence of magnetic flux ropes in the ion diffusion region of
  magnetic reconnection},'' {\em Nature Physics}, vol.~12, no.~3, pp.~263--267,
  2016.

\bibitem{Choi:2016jz}
M.~J. Choi, H.~K. Park, G.~S. Yun, W.~Lee, N.~C. Luhmann, K.~D. Lee, W.~H. Ko,
  Y.~S. Park, B.~H. Park, and Y.~In, ``{2D/3D electron temperature fluctuations
  near explosive MHD instabilities accompanied by minor and major
  disruptions},'' {\em Nuclear Fusion}, vol.~56, no.~6, p.~066013, 2016.

\bibitem{Yun:2014kv}
G.~S. Yun, W.~Lee, M.~J. Choi, J.~Lee, M.~Kim, J.~Leem, Y.~Nam, G.~H. Choe,
  H.~K. Park, H.~Park, D.~S. Woo, K.~W. Kim, C.~W. Domier, N.~C. Luhmann,
  N.~Ito, A.~Mase, and S.~G. Lee, ``{Quasi 3-D ECE Imaging System for Study of
  MHD instabilities in KSTAR},'' {\em Review of Scientific Instruments},
  vol.~85, no.~11, p.~11D820, 2014.

\bibitem{Park:2019kq}
H.~K. Park, ``{Newly uncovered physics of MHD instabilities using 2-D electron
  cyclotron emission imaging system in toroidal plasmas},'' {\em Advances in
  Physics: X}, vol.~4, no.~1, p.~1633956, 2019.

\bibitem{Park:2006il}
H.~K. Park, N.~C. Luhmann, A.~J.~H. Donn{\'e}, I.~G.~J. Classen, C.~W. Domier,
  E.~Mazzucato, T.~Munsat, M.~J. van~de Pol, and Z.~Xia, ``{Observation of
  High-Field-Side Crash and Heat Transfer during Sawtooth Oscillation in
  Magnetically Confined Plasmas},'' {\em Physical Review Letters}, vol.~96,
  no.~19, p.~195003, 2006.

\bibitem{Choi:2019tw}
M.~J. Choi, ``{Spectral data analysis methods for the two-dimensional imaging
  diagnostics},'' {\em https://arXiv.org}, p.~1907.09184v3, 2019.

\end{thebibliography}

\end{document}